\documentclass{ws-p8-50x6-00}

\begin{document}

\def\p{\pi}
\def\ifmath#1{\relax\ifmmode #1\else $#1$\fi}%
\def\rF{\ifmath{{\mathrm{F}}}}
\def\rK{\ifmath{{\mathrm{K}}}}
\def\rp{\ifmath{{\mathrm{p}}}}
\def\rt{\ifmath{{\mathrm{t}}}}
\def\LAB{\ifmath{{\mathrm{LAB}}}}
\def\cut{\ifmath{{\mathrm{cut}}}}
\def\beq{\begin{equation}}
\def\eeq{\end{equation}}

\title{ Event-by-Event Fluctuations of Transverse Momentum
in Elementary Collisions at 250 GeV/$c$}

\author{Bai YuTing, Fu JingHua and Wu YuanFang}

\address{Institute of Particle Physics, HuaZhong Normal University,
WuHan, China\\
E-mail: baiyt@iopp.ccnu.edu.cn}

\maketitle

\abstracts{
We present a study of event-by-event fluctuations of transverse momentum
in $\pi^{+}\rp$ and $\rK^{+}\rp$ collisions at 250 GeV/$c$ and the 
corresponding
PYTHIA Monte Carlo results for $\pi^{+}\rp$ collisions. The dependence of
$\Phi_{p_{\rt}}$ on event and particle variables are investigated
in detail. We find that $\Phi_{p_\rt}$ are all negative for different average 
transverse momentum per event sample. The $\langle p_{\rt} \rangle_N \sim N$ 
correlation may not be the only origin of $\Phi_{p_{\rt}}$.}  

\section{Introduction}
The investigation of single events has attracted  
a lot of attentions in heavy-ion collisions. In which, a method suggested 
by M. Ga$\acute z$dzicki \cite{1} has been frequently used. 
It is defined by
$$\Phi_{p_{\rt}}=\sqrt{\frac{\langle Z^2 \rangle}{\langle N \rangle}}
-\sqrt{\langle z^2 \rangle}$$ 
where $z_{i}=p_{\rt_{i}}-\langle p_{\rt} \rangle$ is a 
single-particle transverse momentum variable,
$Z=\sum\limits_{i=1}^{N} z_{i}$ is an event variable with summation 
running over all $N$ final-state particles in an event, and $\langle 
\cdots \rangle$ denotes the average over all events in the sample.
It was shown that the method removes the trivial fluctuations caused by 
the variation of impact parameter and by statistics, so that the dynamical
event-by-event fluctuations of transverse momentum can be studied. 
The method is also supposed to be sensitive to the correlations 
between multiplicity and average transverse momentum. The larger this 
correlation, the stronger the $\Phi_{p_{\rt}}$.   
It is proposed that vanishing of this correlation can be used as 
a signal of ``equilibration'' of the system created in A-A collisions.

To understand the behavior of $\Phi_{p_\rt}$ in heavy-ion collisions, it is 
important to see how $\Phi_{p_\rt}$ depends on the correlation between average
transverse momentum and multiplicity and how it relates to event and particle 
variables in hadron-hadron collisions.  

We'll show this behavior on $\pi^{+}\rp$ and $\rK^{+}\rp$ collisions
at 250 GeV/$c$ in NA22.
Our analysis will be focused on the central rapidity
range $-2\leq y\leq 2$ 
and best momentum resolution range
0.001 GeV/$c$ $\leq p_{\rt} \leq 10$ GeV/$c$.

\begin{picture}(230,70)
\put(-14,0)
{
      \epsfig{file=fig1.epsi,height=2.3cm}
}
\put(73,0)
{     
      \epsfig{file=fig7.epsi, height=2.3cm}
}
\put(167,0)
{
      \epsfig{file=fig3.epsi, height=2.3cm}
}
\end{picture}
\vskip 0.5mm
\noindent {\scriptsize Figure 1. $\langle p_{\rt}\rangle_{n}\ vs.\ n$ in the 
\hskip 1.mm Figure 2. Depedence of $\Phi_{p_{\rt}}$ 
Figure 3. $\langle p_{\rt}\rangle_{n}\ vs.\ n$ at different $\bar p_{\rt}$ 
\vskip 0.2mm
\noindent  full sample. 
\hskip 2.2cm on $n$.     
\hskip 2.7cm subsamples.} 

\section{$\Phi_{p_{\rt}}$ for the full sample}
The negative correlation between charged $n$ and $\langle p_\rt\rangle_n$
(full trangles) in both the $\pi^{+}\rp$ and $\rK^{+}\rp$ collision data at 250 GeV/$c$ 
is reproduced in Fig.~1, and compared to the expectation (open trangles) 
from PYTHIA \cite{3} 
which underestimates both $\langle p_\rt\rangle_n$ at low $n$ and the $n$ 
dependence of $\langle p_\rt\rangle_n$.

The $\Phi_{p_{\rt}}$  of the full sample is a non-zero positive value of 
$29.06\pm 1.68$ MeV/$c$, while the PYTHIA for $\pi^{+}\rp$ 
collisions is $18.95\pm 0.95$ MeV/$c$, thus underestimating the event-by-event
fluctuation. 

\section{The dependence of $\Phi_{p_{\rt}}$ on event variables}
Now that $\Phi_{p_{\rt}}$ describes the event fluctuations of
transverse momentum, it is interesting to compare the event-by-event 
fluctuation strength in subsamples with different values of $n$ and 
$\bar p_{\rt}={\sum\limits_{i=1}^{n}\rp_{t_{i}}}/n$.

First, we divide the full sample into two multiplicity
subsamples a) $n<\langle n\rangle$ and b) $n\geq\langle n\rangle$.
The $\Phi_{p_{\rt}}$ values for these two
subsamples are given in Table 1. Both of them are positive and the bigger 
multiplicity subsample has larger event-by-event fluctuation. 
For more detail, the dependence of
$\Phi_{p_{\rt}}$ on $n$ is plotted in Fig.~2. $\Phi_{p_{\rt}}$ is always
positive and increases with increasing multiplicity $n$. PYTHIA shows the
same trend, but with lower $\Phi_{p_\rt}$ values.

\vskip 0.2cm
\centerline{\scriptsize Table 1. $\Phi_{p_{\rt}}$ and its related variables 
in the two different $n$ subsamples.}
\begin{center}
\scriptsize
\begin{tabular}{|c|c|c|c|c|c|c|}
\hline
Sample& Number &$\langle p_{\rt} \rangle$& $\langle n \rangle$ &
$\frac{\langle Z^2 \rangle}{\langle n \rangle}$ &
$\langle{z^2}\rangle $ &
{$\Phi_{p_{\rt}}$}  \\
  &of events &GeV/$c$ & & MeV$^2/c^2$ & MeV$^2/c^2$ & MeV/$c$  \\ \hline
a)  & 25844 & 0.396 & 3.99 & 77969.83 & 69680.73
& 15.26$\pm$1.92 \\ \hline
$\pi^{+}\rp$ MC& 50669 & 0.336 & 4.04 & 47307.93 & 43113.39
& 9.87$\pm$1.08  \\ \hline
b) & 17836& 0.380 & 9.99 & 92164.61 & 73047.67
& 33.31$\pm$2.93 \\ \hline
$\pi^{+}\rp$ MC& 45721& 0.328 & 9.34 & 54901.88 & 44999.70
& 22.18$\pm$1.12 \\ \hline
\end{tabular}
\end{center}

\vskip 0.3cm
Then, the full sample is separated into two according to
a) $\bar p_{\rt}< \langle p_{\rt} \rangle$ and b)
$\bar p_{\rt}\geq \langle p_{\rt} \rangle$. Their corresponding 
$\langle p_{\rt}\rangle_{n}\ vs.\ n$ correlation is plotted in Fig.~3.

\begin{picture}(230,70)
\put(-4,0)
{
      \epsfig{file=fig5.epsi,height=2.5cm}
}
\put(180,0)
{
      \epsfig{file=fig9.epsi,height=2.5cm}
}
\end{picture}
\vskip 0.2mm
\noindent {\scriptsize Figure 4. Depedence of 
$\Phi_{p_{\rt}}$ on $\bar p_{\rt}$. 
\hskip 1.5cm 
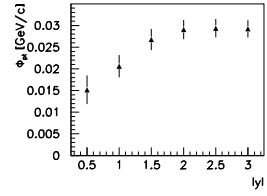
Figure 5. Dependence of $\Phi_{p_{\rt}}$
on rapidity range.}

\vskip 0.2cm
\noindent
We can see a positive correlation for
subsample a) 
first increasing with multiplicity and then saturating.
For subsample b), there 
is a negative correlation between $\langle p_{\rt} \rangle_n$ and $n$.

The values of $\Phi_{p_{\rt}}$ for these two subsamples are given in Table 2.
It is astonishing that they are both negative. 
The corresponding Monte Carlo results are provided in the same table 
and figure. Qualitatively, they show the same correlation tendencies as 
the experimental data, but quantitatively they underestimate the 
experimental ones. We wonder if $\Phi_{p_\rt}$ can still present the 
event-by-event fluctuation in this case.

\vskip 0.2cm
\centerline{\scriptsize Table 2. $\Phi_{p_{\rt}}$ and its related variables for the two different $\bar p_{\rt}$ samples.}
\begin{center}
\scriptsize
\begin{tabular}{|c|c|c|c|c|c|c|}
\hline
Sample& Numer &$\langle p_{\rt} \rangle$& $\langle n \rangle$ &
$\frac{\langle Z^2 \rangle}{\langle n \rangle}$ &
$\langle {z^2}\rangle$ &
{$\Phi_{p_{\rt}}$}  \\
  &of events &GeV/$c$ & & MeV$^2/c^2$ & MeV$^2/c^2$ & MeV/$c$  \\ \hline
a) & 22844 & 0.309 & 7.18 & 21511.77 & 38514.35 & -49.58$\pm$1.67
\\ \hline
$\pi^{+}\rp$ MC& 51960& 0.270 & 6.65 &13410.97 & 26317.27 & -46.42$\pm$0.59
\\ \hline
b) & 20836 & 0.478 & 6.72 & 54062.51 & 98232.83 & -80.91$\pm$3.08  
\\ \hline
$\pi^{+}\rp$ MC& 44430& 0.404 & 6.44 & 30923.91 & 56424.77 & -61.69$\pm$1.19
\\ \hline
\end{tabular}
\end{center}

\vskip 0.3cm
For more detail, $\Phi_{p_\rt}$ for different $\bar p_\rt$ subsamples is given
in Fig.~4. We can see that $\Phi_{p_\rt}$ is negative for any $\bar p_\rt$ 
subsample and decreases monotonously with increasing $\bar p_\rt$. 
PYTHIA gives the same trend, but with slightly less negative values.

\section{The dependence of $\Phi_{p_{\rt}}$ on the particle variable}
Now we turn to discuss how $\Phi_{p_\rt}$ behaves for different transverse 
momentum particles. Four samples are defined by the cut $p_\rt > p_\rt^{\cut}$
with $p_\rt^{\cut}=0.1, 0.2, 0.3$ and 0.4 GeV/$c$. 
It was shown in \cite{2}   
that the lowest-$p_{\rt}^{\cut}$ subsample has the strongest negative 
correlation and the highest-$p_{\rt}^{\cut}$ sample has the strongest positive 
one.  

In Table 3, the $\Phi_{p_{\rt}}$ values are listed.
Here we observe, the higher the $p_{\rt}^{\cut}$ and stronger the correlation, 
the smaller $\Phi_{p_{\rt}}$. This is in contradiction with arguments that 
$\Phi_{p_\rt}$ are mainly due to $\langle p_{\rt}\rangle_{n}\ vs.\ n$ 
correlations ({\it cf.} Fig.4 of \cite{1}).

\newpage
\centerline{\scriptsize Table 3. $\Phi_{p_{\rt}}$ and its related variables for
different $p_{\rt}^\cut$ subsamples.}
\vskip 0.2cm
\begin{center}
\scriptsize
\begin{tabular}{|c|c|c|c|c|c|c|}
\hline
Sample & Number &$\langle p_\rt \rangle$&
$\langle N \rangle$ &
$\frac{\langle Z^2 \rangle}{\langle N \rangle}$ & $\langle{z^2}\rangle $ &
{$\Phi_{p_{\rt}}$} \\
        & of events  & GeV/$c$ & & MeV$^2/c^2$ & MeV$^2/c^2$& MeV/$c$ \\ \hline
 $p_{\rt}^\cut>0.1$GeV/$c$& 43489 & 0.412 & 6.45 &  
82967.44 & 68776.52 &25.79$\pm$2.18 \\  \hline
 $\pi^{+}\rp$ MC & 95937& 0.359 & 5.96 & 47871.61 &
40685.00 & 17.09$\pm$0.84 \\ \hline
 $p_{\rt}^\cut>0.2$GeV/$c$& 42912& 0.474 & 5.25 & 
76051.82 & 65047.70 &20.73$\pm$2.14 \\ \hline 
$\pi^{+}\rp$ MC & 94390 & 0.420 & 4.68 & 43234.06 & 
36107.48 & 17.91$\pm$0.87\\ \hline
 $p_{\rt}^\cut>0.3$GeV/$c$& 41505 & 0.557 & 3.95 & 
71900.70 & 63178.21 & 16.79$\pm$2.40 \\ \hline
 $\pi^{+}\rp$ MC & 90394& 0.497 & 3.36 & 39290.56 & 
32841.53 & 17.00$\pm$0.97\\ \hline
 $p_{\rt}^\cut>0.4$GeV/$c$& 38583 & 0.649 & 2.92 & 
70183.20 & 62918.54 &14.09$\pm$2.85 \\ \hline 
 $\pi^{+}\rp$ MC & 80820& 0.583 & 2.39 & 36320.72 & 
30920.49 & 14.73$\pm$1.14 \\ \hline
\end{tabular}
\end{center}

\section{The dependence of $\Phi_{p_{\rt}}$ on rapidity range}
It is important whether the rapidity range used for the analysis influences 
the $\Phi_{p_{\rt}}$. So, $\Phi_{p_\rt}$ is given in Fig.~5 for different 
central rapidity ranges. When the rapidity range broadens from the center, 
$\Phi_{p_{\rt}}$ becomes larger and larger. Once the range spreads to wider 
than $-2\leq y \leq 2$, $\Phi_{p_{\rt}}$ reaches its saturation value. 
So, this result confirms that the $\Phi_{p_{\rt}}$ we obtained from the 
central rapidity range $-2\leq y \leq 2$,  
can represent the full one.

\section{Summary}
\hskip 6mm
1. $\Phi_{p_\rt}$ is positive both for the full sample and for all samples of 
restricted multiplicity, and the high-multiplicity event sample has the  
strong event-by-event fluctuation. But it is negative for  
all samples of restricted $\bar p_{\rt}$.  

2. The higher the $p_\rt$ in the sample, the stronger 
$\langle p_{\rt}\rangle_{n}\ vs.\ n$ correlation, but  
the weaker $\Phi_{p_\rt}$. It turns out that this
correlation may not be the only origin of event-by-event fluctuation.


3. Except for samples with particle transverse momentum larger than the 
average, the $|\Phi_{p_\rt}|$ values are underestimated by PYTHIA. 

\section*{Acknowledgments}
We would like to thank Prof. Liu Feng for the important suggestions.


\vskip -5mm
\begin{thebibliography}{99}
\bibitem{1} M. Ga$\acute z$dzicki, St. Mr\'owczy\'nski, Z. Phys.
C54 (1992) 127. 

\bibitem{2} V.V. Aivazyan et al. (NA22); Phys. Lett. B209 (1988) 103.

\bibitem{3} T. Sjostrand, Computer Physics Commun. 82 (1994) 74. 

\end{thebibliography}
\end{document}